\documentclass[12pt]{article}
\usepackage{epsfig}
\usepackage{subfigure}
\usepackage{url}

\graphicspath{{./figures/}}
\def\beq{\begin{equation}}
\def\eeq#1{\label{#1}\end{equation}}
\def\eeqn{\end{equation}}

\def\beqa{\begin{eqnarray}}
\def\eeqa#1{\label{#1}\end{eqnarray}}
\def\eeqan{\end{eqnarray}}

\let\bar=\overbar

\def\Dslash{\not{\hbox{\kern-4pt $D$}}}
\def\dslash{\not{\hbox{\kern-2pt $\del$}}}

\def\BR{\mbox{\rm BR}}
\def\SES{\mbox{\rm SES}}

\def\msb{{\bar{\ssstyle M \kern -1pt S}}}

\def\Title#1{\begin{center} {\Large {\bf #1} } \end{center}}

\begin{document}

%% !!! Bakul Gaur: these lines can be safely removed in the end !!!
% \linenumbers
% Abstract: The rare decays $B_{s,d}^0 \to \mu^+\mu^-$ are important probes for new physics beyond the Standard Model. These decays are subjects of studies at the LHC. The ATLAS and CMS collaborations have analyzed the proton-proton collisions data at 7~TeV centre-of-mass energy collected in the year 2011. They have established upper limits on the branching fractions of the two decays. This paper presents a report.

\Title{Search for Pure Leptonic B Decays at ATLAS and CMS}

\bigskip\bigskip

%+\addtocontents{toc}{{\it D. Reggiano}}
%+\label{ReggianoStart}

\begin{raggedright}  

{\it Bakul Gaur\index{Gaur, B.} on behalf of the ATLAS and CMS Collaborations\footnote{This work was supported in part by BMBF.} \\
Naturwissenschaftlich-Technische Fakult\"{a}t - Department f\"{u}r Physik\\
Universit\"{a}t Siegen\\
D-57068 Siegen, GERMANY}
\bigskip\bigskip
\end{raggedright}

\begin{raggedright}
{\it Proceedings of CKM 2012, the $7^{th}$ International Workshop on the CKM Unitarity Triangle, University of Cincinnati, USA, 28 September - 2 October 2012}
\bigskip\bigskip
\end{raggedright}

\section{Introduction}

In the Standard Model (SM) the decays $B_{s,d}^0 \to \mu^+\mu^-$ are mediated by flavour changing neutral currents (FCNC). They are CKM suppressed, and happen via higher-order electroweak Feynman diagrams. Their decay rates, additionally, suffer helicity suppression. For the $B_{s}^0 \to \mu^+\mu^-$ channel a rate of $(3.54\pm0.30)\times10^{-9}$ has been predicted \cite{Buras2012,DeBruyn2012}, and for $B_{d}^0 \to \mu^+\mu^-$ the prediction is $(1.07\pm0.10)\times10^{-10}$ \cite{Buras2012}. In Standard Model extensions these rates may be enhanced. Any deviation from the SM predictions will indicate new physics. These channels are important probes for physics beyond the SM.

This paper describes the latest measurements\footnote{as of summer 2012} of $B_{s,d}^0 \to \mu^+\mu^-$ from the ATLAS and CMS experiments \cite{ATLAS,CMS} at the LHC. %The next section introduces the general analysis strategy. Sections~\ref{sec:bsmumu_atlas} and \ref{sec:bsmumu_cms} summarize the analyses of ATLAS and CMS.

\section{Analysis Strategy}
\label{sec:bsmumu_strategy}

The branching fractions $\BR(B_{s,d}^0 \to \mu^+\mu^-)$ can be measured relative to a well measured reference channel to minimize uncertainties, e.g. detector acceptance uncertainties, while keeping the analysis independent of luminosity variations and $b\bar{b}$ production cross-section uncertainties. For this purpose both ATLAS and CMS use $B^\pm \to J/\psi K^\pm \to \mu^+\mu^- K^\pm$ as the reference channel, because of large available statistics, and similar (di-muon) final states. To keep systematic uncertainties low, similar selection cuts are applied to both the signal and the reference channel events.

The branching fractions are computed as
\begin{equation}
\label{eqn:bsmumu_br}
\BR(B_{s,d}^0 \to \mu^+\mu^-)=\frac{N_{\mu^+\mu^-}}{N_{J/\psi K^\pm}}\times\BR(B^\pm \to J/\psi K^\pm \to \mu^+\mu^- K^\pm)\times R_{A\epsilon}\times\frac{f_u}{f_{s,d}},
\end{equation}
where $N_{\mu^+\mu^-}$ and $N_{J/\psi K^\pm}$ are the number of observed signal and reference channel events, respectively. The factor $R_{A\epsilon}=\frac{A_{J/\psi K^\pm}\epsilon_{J/\psi K^\pm}}{A_{\mu^+\mu^-}\epsilon_{\mu^+\mu^-}}$ is to correct for the detector acceptances ($A$) and event selection efficiencies ($\epsilon$) estimated for the two channels using Monte Carlo events (MC). The ratios $f_u/f_{s,d}$ are the ratios of the $b$-quark hadronization probabilities to correct for the different production rates of $B^\pm$ and $B_{s,d}^0$. The reference channel branching fraction $\BR(B^\pm \to J/\psi K^\pm \to \mu^+\mu^- K^\pm)=(6.01\pm0.21)\times10^{-5}$, and the ratio $f_{s}/f_u=0.267\pm0.021$ are taken from other measurements \cite{PDG2010,FuFs2012}. The ratio $f_d/f_u$ is taken to be 1 \cite{Asner2010}.

\section{ATLAS Analysis}
\label{sec:bsmumu_atlas}

The ATLAS analysis expresses the branching fraction as a product of the observed number of signal events and a Single Event Sensitivity (SES):
\begin{equation}
\label{eqn:bsmumu_ses}
\BR(B_{s}^0 \to \mu^+\mu^-)=N_{\mu^+\mu^-}\times\SES.
\end{equation}
For a single observed signal event, the branching fraction $\BR(B_{s}^0 \to \mu^+\mu^-)$ would be equal to the SES.

The analysis uses $pp$ collision data at $\sqrt{s}=7$~TeV recorded by the ATLAS detector in the period April-August 2011. This corresponds to an integrated luminosity of 2.4~fb$^{-1}$. The analysis is robust against pileup effects. The details of the analysis can be found in reference \cite{BsMuMuATLAS2012}.

A topological trigger selects di-muon candidates above a transverse momentum ($p_T$) threshold of 4~GeV. The signal channel events in the invariant mass range $m_{\mu^+\mu^-}\in[5066,5666]$~MeV are hidden in the analysis ('blind' analysis) until event selection cuts are finalized. The events in the side bands ($m_{\mu^+\mu^-}\in[4766,5066]$~MeV, $m_{\mu^+\mu^-}\in[5666,5966]$~MeV) are split into two sets. To avoid statistical biases, one set is used for cut optimization (odd numbered events in data), and the other for estimation of the background under the signal (even numbered events in data).

The event selection cuts are optimized using a multivariate technique. The method uses Boosted Decision Trees (BDT) with 14 input variables to compute an event classifier, $Q$.
% The BDT classifier distributions for the signal MC sample and the side-band events (odd events) are shown in Figure~\ref{fig:atlas_bdt}. The figure also shows a comparison of the classifier distributions of the reference channel MC sample and the side-band-subtracted\footnote{The side bands in the reference channel are defined as $m_{J/\psi K^\pm}\in[4930,5130]$~MeV, $m_{J/\psi K^\pm}\in[5430,5630]$~MeV.} reference channel data.
The signal search window, $\Delta m_{\mu^+\mu^-}$, in the $B_s^0 \to \mu^+\mu^-$ invariant mass spectrum is optimized together with the BDT classifier $Q$. The method determines the optimal $\Delta m_{\mu^+\mu^-}$ and $Q$ to get the maximum value for the estimator:
\begin{equation}
\label{eqn:punzi_estimator}
\mathcal{P}(Q,\Delta m_{\mu^+\mu^-})=\frac{\epsilon_{sig}}{1+\sqrt{N_{bkg}}},
\end{equation}
where $\epsilon_{sig}$ is the signal selection efficiency and $N_{bkg}$ is the continuum background interpolated from the side bands (two times the number of odd events).

% \begin{figure}[htb]
% \begin{center}
% \vspace{0.08in}
% \epsfig{file=atlas_bdt.eps,hedight=2.5in}
% \caption{BDT event classifier ($Q$) distributions. Left: a comparison of the $Q$-distributions for $B_s^0 \to \mu^+\mu^-$ MC sample (squares) and the side-band events in ATLAS data (points). Right: a comparison of the $Q$-distribution for the $B^\pm \to J/\psi K^\pm \to \mu^+\mu^- K^\pm$ MC sample (triangles) and the side-band-subtracted data (stars).}
% \label{fig:atlas_bdt}
% \end{center}
% \end{figure}

The events are split into three mass resolution categories distinguished by the maximum pseudorapidity ($|\eta^\mu|_{max}$) of the muon tracks. The $\Delta m_{\mu^+\mu^-}$ and $Q$ are separately optimized for each category. The same classifier cut is used to compute the acceptance and efficiency ratio $R_{A\epsilon}$ from the MC, and the reference channel yield, $N_{J/\psi K^\pm}$, in the three categories. The $B^\pm$ yield is determined by fitting the $B^\pm \to J/\psi K^\pm \to \mu^+\mu^- K^\pm$ invariant mass spectrum, and computing the $N_{J/\psi K^\pm}$ in a band $m_{J/\psi K^\pm}\in[5180,5380]$~MeV. The SES is thus computed for the three categories using Equations~(\ref{eqn:bsmumu_br}) and (\ref{eqn:bsmumu_ses}). In the di-muon invariant mass spectrum there is an expected background due hadronic decays, where hadrons are misidentified as muons. The $B \to hh$ background is the dominating resonant background. It is estimated from the MC.

\begin{figure}[htb]
\begin{center}
\vspace{0.08in}
\subfigure{\epsfig{file=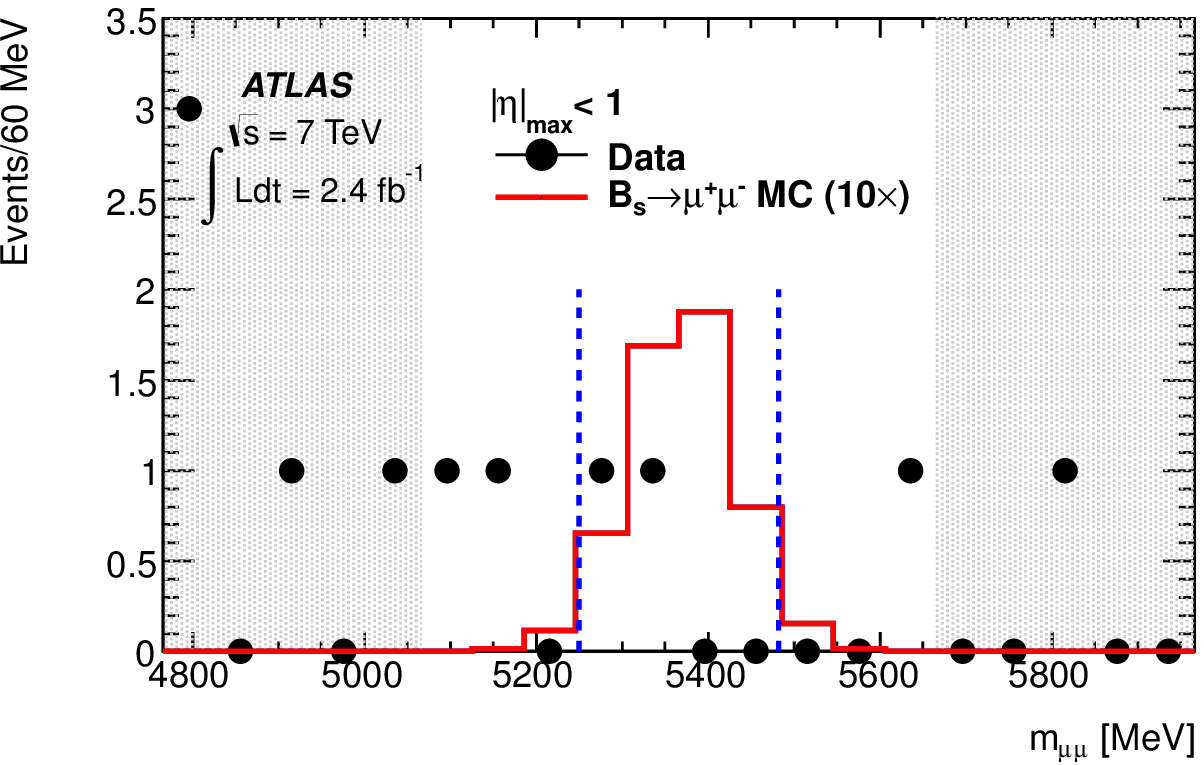,height=1.83in}}\hspace{0.12in}
\subfigure{\epsfig{file=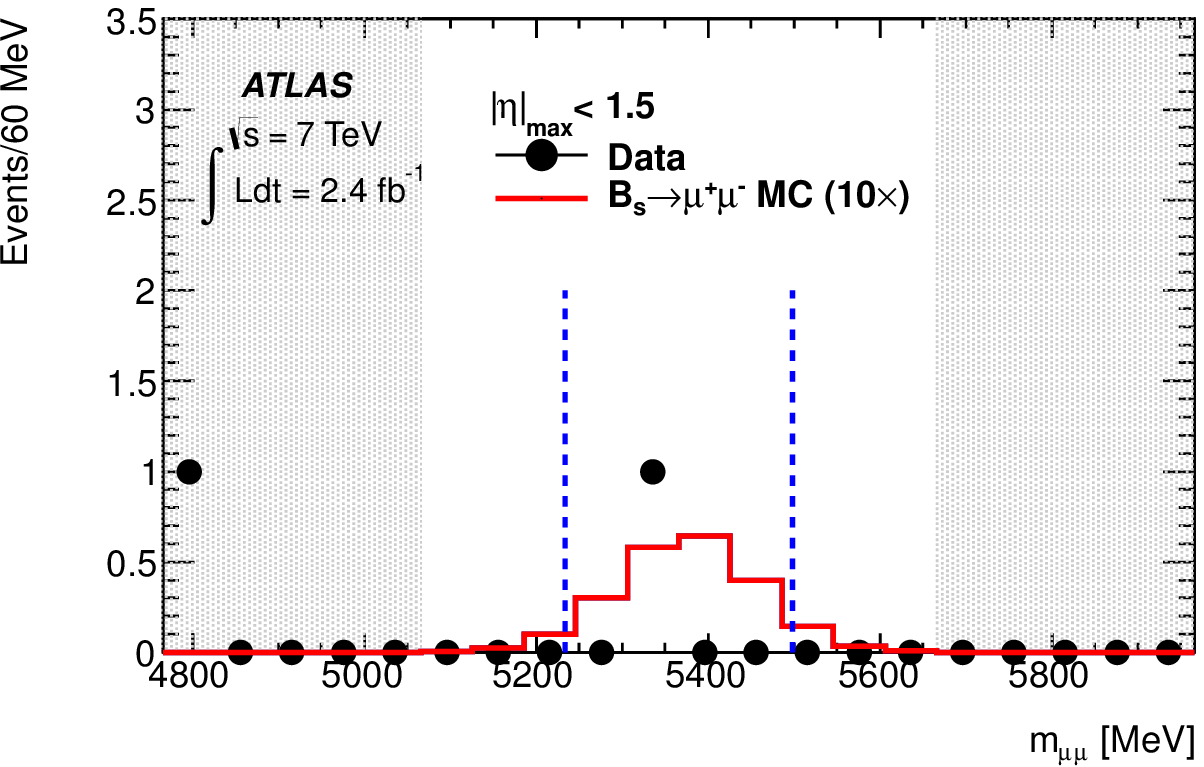,height=1.83in}}\vspace{0.04in}
\subfigure{\epsfig{file=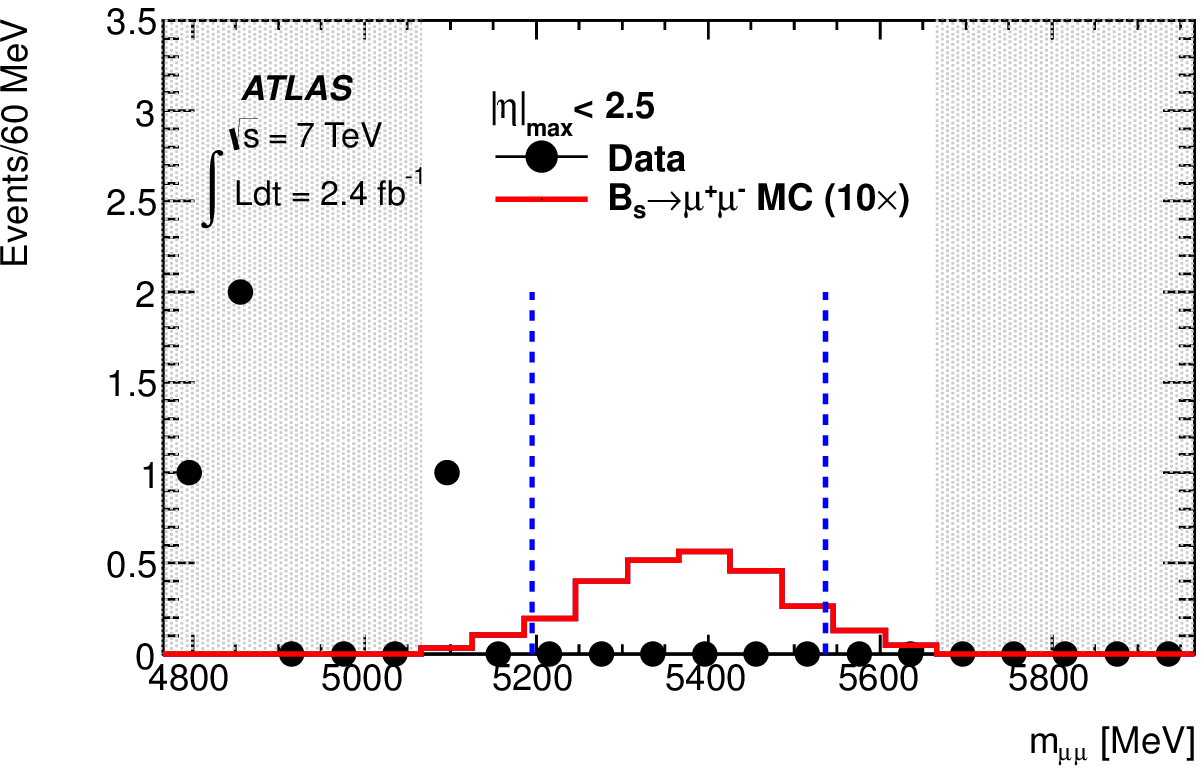,height=1.83in}}
\caption{The di-muon invariant mass spectrum (points) (in the ATLAS analysis \cite{BsMuMuATLAS2012}) after unblinding in the three mass resolution categories. The search window is shown by dashed blue lines. The shaded grey regions reflect the sidebands used in the analysis. The solid curve shows the $B_{s}^0 \to \mu^+\mu^-$ distribution in MC scaled by a factor of 10.}
\label{fig:atlas_signal}
\end{center}
\end{figure}

\subsection{Results}
\label{sec:bsmumu_atlas_res}

\begin{figure}[htb]
\begin{center}
\vspace{0.08in}
\epsfig{file=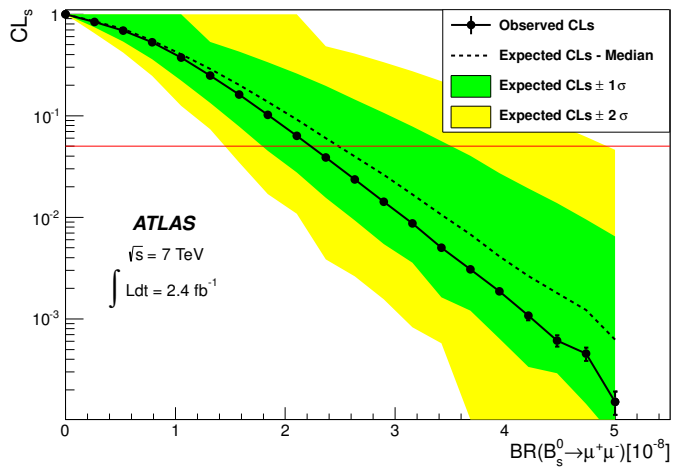,height=2.4in}
\caption{The observed CL$_\mathrm s$ (points) as a function of $B_s^0 \to \mu^+\mu^-$ branching fraction in the ATLAS analysis \cite{BsMuMuATLAS2012}. The upper limit is read at the intersection of the red line with the observed CL$_\mathrm s$ line, which corresponds to 95\% CL. The green and yellow bands indicate $\pm1\sigma$ and $\pm2\sigma$ deviation from the expected limit (dashed line).}
\label{fig:atlas_cls}
\end{center}
\end{figure}

The upper limit on the $B_{s}^0 \to \mu^+\mu^-$ branching fraction is computed using the CL$_\mathrm s$ method \cite{Read2002}. A likelihood expression combines the SES computed in the three mass resolution categories, its uncertainties, and the expected resonant ($B \to hh$) and non-resonant background contributions to the $B_{s}^0 \to \mu^+\mu^-$ invariant mass spectrum \cite{BsMuMuATLAS2012}. An expected limit of $(2.3^{+1.0}_{-0.5})\times10^{-8}$ on $\BR(B_s^0 \to \mu^+\mu^-)$ is obtained. The data in the blinded region is analysed, and the signal yield is measured in the optimized search window for the three mass resolution categories (see Figure~\ref{fig:atlas_signal}). The observed upper limit is $2.2\ (1.9)\times10^{-8}$ at 95\% (90\%) confidence level (CL). The observed limit is comparable with the expected limit (Figure~\ref{fig:atlas_cls}). %The observed CL$_\mathrm s$ distribution as a function of the $B_{s}^0 \to \mu^+\mu^-$ branching fraction is shown in Figure~\ref{fig:atlas_cls}.

\section{CMS Analysis}
\label{sec:bsmumu_cms}

In this section the key features in the CMS analysis different from the ATLAS one are presented.

The CMS analysis uses 5~fb$^{-1}$ of $pp$ collisions at $\sqrt{s}=7$~TeV recorded by the CMS detector in the year 2011. In this analysis a cut-and-count approach is taken. The optimization of the selection cuts is performed using the signal MC and all the side-band ($m_{\mu^+\mu^-}\in[4900,5200]$~MeV, $m_{\mu^+\mu^-}\in[5450,5900]$~MeV) events in the data. The events inside the signal region (in data) are kept blinded until the cuts are established. A random-grid search method is used to tune the cuts on 11 analysis variables to get the best upper limit \cite{BsMuMuCMS2012}. A different set of selection cuts is used for events in the endcap region of the detector than in the barrel region.

The analysis shows good agreement between the reconstructed distributions in data and MC, and it is not sensitive to pileup events.

\subsection{Results}
\label{sec:bsmumu_cms_res}

\begin{figure}[htb]
\begin{center}
\vspace{0.08in}
\subfigure{\epsfig{file=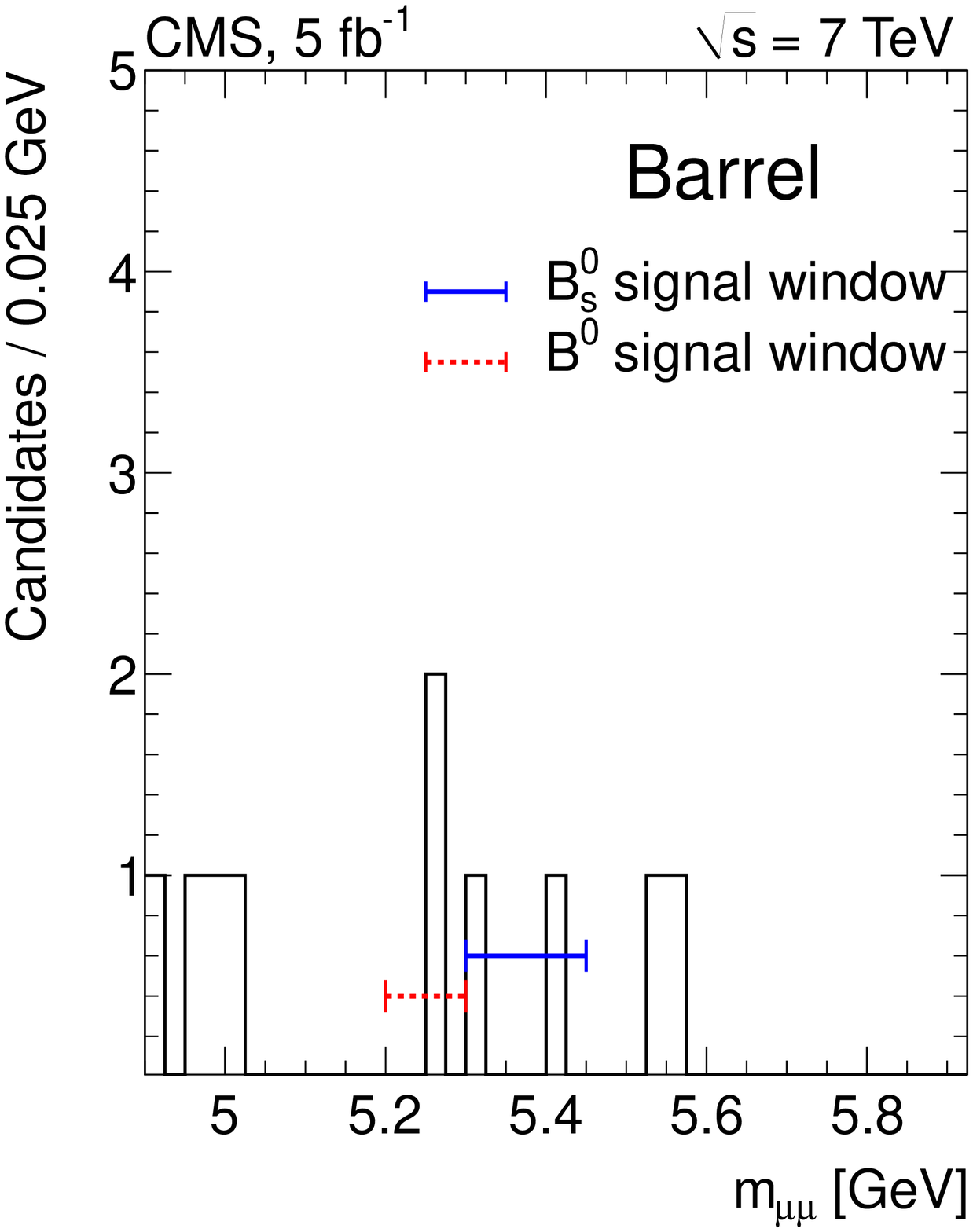,height=2.5in}}
\subfigure{\epsfig{file=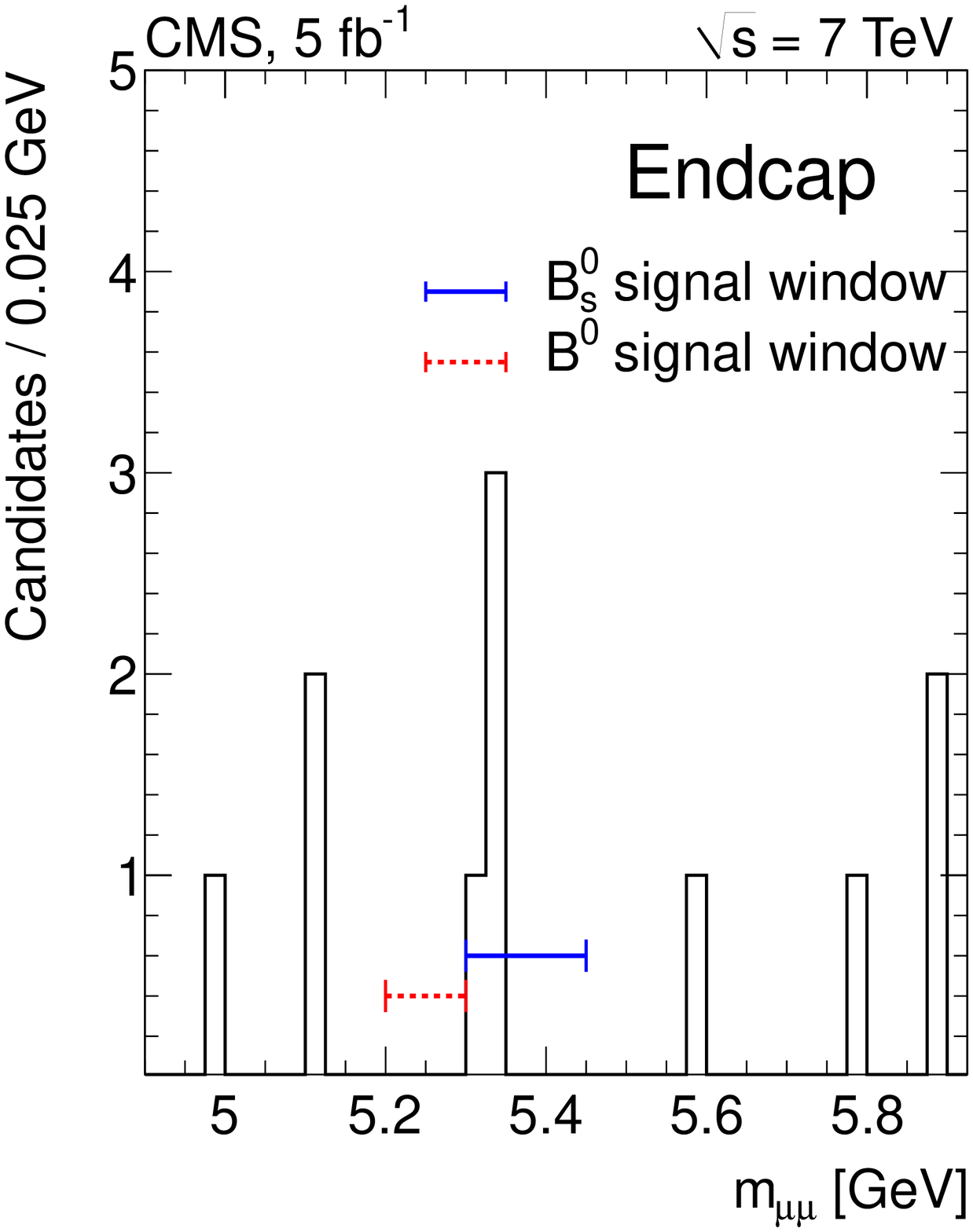,height=2.5in}}
\caption{The di-muon invariant mass spectrum (in the CMS analysis \cite{BsMuMuCMS2012}) after unblinding in the two detector regions. The solid and the dashed horizontal lines indicates the search windows for $B_{s}^0$ and $B_{d}^0$ events, respectively.}
\label{fig:cms_signal}
\end{center}
\end{figure}

The branching fractions for both $B_{d}^0 \to \mu^+\mu^-$ and $B_{s}^0 \to \mu^+\mu^-$ decays are measured simultaneously using two asymmetric search windows around the $B_{d}^0$ and $B_{s}^0$ masses. Figure~\ref{fig:cms_signal} shows the di-muon invariant mass spectrum after the unblinding.

\begin{figure}[htb]
\begin{center}
\vspace{0.08in}
\epsfig{file=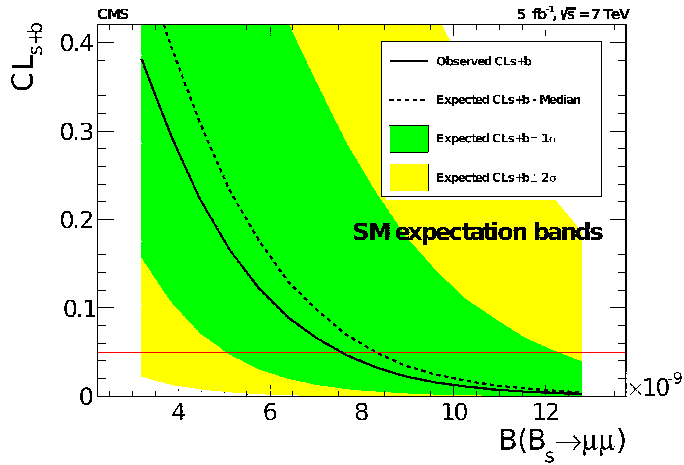,height=2.4in}
\caption{The observed CL$_\mathrm {s+b}$ (solid line) as a function of $B_s^0 \to \mu^+\mu^-$ branching fraction (in the CMS analysis \cite{BsMuMuCMS2012}) assuming the SM expectations. The green and yellow bands indicate $\pm1\sigma$ and $\pm2\sigma$ deviation from the expected limit (dashed line).}
\label{fig:cms_clsb}
\end{center}
\end{figure}

The limit extraction takes into account the expected combinatorial and the resonant background contributions (estimated using MC), and also the number of expected signal events assuming the SM branching fractions. The resonant background includes $B_{s,d}^0 \to h^+h^{(')-}$ decays, where hadrons are misidentified as muons, and $B_{s,d}^0 \to h^-\mu^+\nu$ rare semileptonic decays. The limits are computed using the CL$_\mathrm s$ method \cite{Read2002,Junk1999}. The expected upper limits for $B_{s}^0 \to \mu^+\mu^-$ ($B_{d}^0 \to \mu^+\mu^-$) are $8.4\times10^{-9}$ ($1.6\times10^{-9}$) at 95\% CL \cite{BsMuMuCMS2012}. The upper limit for $B_{d}^0 \to \mu^+\mu^-$ is $1.8\ (1.4)\times10^{-9}$ at 95\% (90\%) CL. Figure~\ref{fig:cms_clsb} shows the dependence of CL$_\mathrm {s+b}$ on the $B_{s}^0 \to \mu^+\mu^-$ branching fraction.

\section{Summary}
\label{sec:summary}

The $B_{s}^0 \to \mu^+\mu^-$ and $B_{d}^0 \to \mu^+\mu^-$ are important decay channels to search for new physics. The latest limits on their branching fractions were set by the LHC experiments. These are summarized in Table~\ref{tab:bsdmumu_limits}. The best $B_{s}^0 \to \mu^+\mu^-$ upper limit, $4.2 \times 10^{-9}$, is the combined result \cite{BsMuMuLHCCombined2012} of the measurements performed by ATLAS, CMS and LHCb experiments. It is close to the SM expectation, $(3.54\pm0.30)\times10^{-9}$. The ATLAS and CMS collaborations are working on extending their analyses to the full available statistics collected from the LHC.

\begin{table}[b]
\begin{center}
\begin{tabular}{l|ccc}  
             & $B_s^0 \to \mu^+\mu^-$ & $B_d^0 \to \mu^+\mu^-$ \\ \hline
ATLAS        & $2.2 \times 10^{-8}$   & --                     \\
CMS          & $7.7 \times 10^{-9}$   & $1.8 \times 10^{-9\ }$ \\
LHCb         & $4.5 \times 10^{-9}$   & $1.0 \times 10^{-9\ }$ \\
LHC Combined & $4.2 \times 10^{-9}$   & $8.1 \times 10^{-10}$  \\ \hline
\end{tabular}
\caption{Limits at 95\% CL measured by different LHC experiments \cite{BsMuMuATLAS2012,BsMuMuCMS2012,BsMuMuLHCb2012,BsMuMuLHCCombined2012}.}
\label{tab:bsdmumu_limits}
\end{center}
\end{table}

\end{document}